\documentclass{elsart}
\usepackage{graphicx}
\usepackage{amssymb}

\newcommand{\noko}{Nokogiriyama underground cell}
\newcommand{\lab}{laboratory}

\newcommand{\kam}{Kamioka Observatory}

\newcommand{\dil}{dilution refrigerator} 
 
\newcommand{\HPGe}{low-background Ge spectrometer}
\begin{document}
\intextsep 30pt
\textfloatsep 30pt
\begin{frontmatter}
\title{First results from dark matter search experiment with LiF bolometer at Kamioka Underground Laboratory}

\author[1]{\corauthref{cor1}\thanksref{0}K.~Miuchi},
\ead{miuchi@icepp.s.u-tokyo.ac.jp}
\thanks[0]{{\it Present Address:} Cosmic-Ray Group, Department of Physics, Faculty of Science, Kyoto University Kitashirakawa, Sakyo-ku, Kyoto, 606-8502, Japan}
\author[1,2]{M.~Minowa},
\author[1]{A.~Takeda},
\author[1]{H.~Sekiya},
\author[1]{Y.~Shimizu},
\author[2,3]{Y.~Inoue},
\author[3]{W.~Ootani},
\author[4]{Y.~Ootuka}
\address[1]{Department of Physics, School of Science, University of Tokyo, 7-3-1, Hongo, Bunkyo-ku, Tokyo 113-0033, Japan}
\address[2]{Research Center for the Early Universe(RESCEU), School of Science, University of Tokyo, 7-3-1, Hongo, Bunkyo-ku, Tokyo 113-0033, Japan}
\address[3]{International Center for Elementary Particle Physics(ICEPP), University of Tokyo, 7-3-1, Hongo, Bunkyo-ku, Tokyo 113-0033, Japan}
\address[4]{Institute of Physics, University of Tsukuba, 1-1-1 Ten'nodai, 
Tsukuba, Ibaraki 305-8571, Japan}
\corauth[cor1]{Corresponding author.}
\begin{abstract}
The Tokyo group has performed the first underground dark matter search
experiment from 2001 through 2002 at {\kam}(2700m.w.e). The detector
is eight LiF bolometers with a total mass of 168g and aims for the direct detection of WIMPs via spin-dependent interaction. With an exposure of 4.1 kg$\cdot$days, we derived the limits in the $a_{\rm p}-a_{\rm n}$ (WIMP-nucleon couplings) plane and excluded a large part of the parameter space allowed by the UKDMC experiment.

\end{abstract}


\end{frontmatter}

\newpage
\section{Introduction}
\label{intro}
Weakly Interacting Massive Particles (WIMPs) are thought to be the
most promising dark matter candidates and many experiments have been
performed for direct detection of WIMPs. The DAMA experiment reported
having detected WIMPs by an annual modulation signal with their 10kg
NaI detectors\cite{DAMA annual}. The reported WIMP parameter region is
$M_{\rm WIMP}=(52^{+10}_{-8})\rm GeV$ and $\rm \sigma^{\rm
SI}_{\rm WIMP-p}=(7.2^{+0.4}_{-0.9})\cdot10^{-6}pb$. On the other
hand, the CDMS Collaboration\cite{CDMS 2000} and EDELWEISS
Collaboration\cite{EDELWEISS 2001} reported having excluded most of
the parameter space claimed by the DAMA experiment within the standard
supersymmetry(scalar couplings) and the halo models. These experiments
aim to detect WIMPs via the spin-independent interactions. In
2001, the DAMA group published a paper in which they interpreted the
annual modulation signal in a spin-independent and spin-dependent
mixed framework\cite{DAMA_SDSI}. According to Ref. \cite{DAMA_SDSI},
$\rm \sigma^{\rm SD}_{\rm WIMP-p}$ could be as high as 0.6pb for $30
\sim 80 \rm GeV$ if we assume certain sets of parameters. These
results suggest that WIMPs could be detected via spin-dependent
interaction and the dark matter search experiments should be performed both via spin-independent and spin-dependent interactions.

We, the Tokyo group, had developed a lithium fluoride (LiF) bolometer
to detect WIMPs via spin-dependent interaction in the surface
laboratory\cite{Tokyo surface} then performed pilot runs at the shallow depth site\cite{Tokyo_1999_PLB}. We have improved our LiF bolometer since we installed the detector system in 2000 at {\kam} (2700m.w.e) and recently we made a dark matter search measurement from 2001 through 2002.

In this paper, we report our first results from the dark matter search experiment performed at {\kam} with new limits in the $a_{\rm p}-a_{\rm n}$ (WIMPs-nucleon couplings) plane.

\section{Experimental set-up and measurements}
\label{set-up}
The laboratory for this dark matter search experiment was newly caved
near the Super-Kamiokande detector at Mozumi Mine of Kamioka Mining
and Smelting Co. located at Kamioka-cho, Gifu, Japan. The {\lab} is at
 the depths of 2700 m.w.e and the muon flux is reduced about five orders of
magnitude compared to the surface \lab. As the radon concentration in
the mine air is high and has a seasonal change (40 (winter)$\sim$1200
(summer) Bq/$\rm m^3$), the fresh air  from the outside of the mine is blown into the {\lab} with a rate of 1 $\rm m^3/min$. The radon concentration in the {\lab} is less than 50 Bq/$\rm m^3$.

The detector used in this dark matter search experiment is a LiF bolometer array consisting of eight LiF bolometers with a total mass of 168g. Each bolometer consists of a 2$\times$2$\times$2 cm${}^3$  LiF crystal and a neutron transmutation doped (NTD) Ge thermistor (1.5$\times$1$\times$1 mm${}^3$ )\cite{NTD} glued to the LiF  with GE varnish. We refer to each bolometer as D1, D2,..., and D8. 
Four of the LiF crystals are purchased from BICRON Co. Ltd. (expressed
as BIC in Table \ref{boloIDs}) while the others are from OHYO KOKEN KOGYO Co. Ltd. (expressed as OKEN in Table \ref{boloIDs}).
NTD3 thermistors, which have been used for most of the previous
measurements, are used for six out of the eight bolometers, while NTD2 and NTD4 are used for one bolometer each. Characteristics of the NTD thermistors are shown in Ref. \cite{NTD}. 
Radioactive contamination in the crystals were previously checked with a {\HPGe} and was found to be less than 0.2 ppb for U, 2 ppb for Th, and 20 ppm for K. Surfaces of the LiF crystals were etched with perchloric acid in order to eliminate remaining grains of the polishing powder. 
Features of the eight bolometers are summarized in Table \ref{boloIDs}.

Schematic drawing of the bolometer array is shown in Fig. \ref{sche
bolometer array}. The array consists of two stages, each stage holding
four bolometers. The bolometers are fixed with spring pins and Delrin
balls. Thermal contacts between the LiF crystals and the oxygen free
high conductivity (OFHC) copper holder are achieved by annealed OFHC
copper ribbons of 10 mm width and 0.1 mm thickness. The etching
process and the assembly were performed in a clean bench of class
100. The bolometer array is encapsulated in a low-radioactivity lead
(older than 200 years) of 2cm thickness.  The OFHC copper holder and
the inner shield were etched with nitric acid in order to eliminate the
surface contamination.

\begin{table}[h]
\begin{center}
\begin{tabular}{
 l|cccccc cc}
\hline
Detector&D1&D2&D3&D4&D5&D6&D7&D8 \\ \hline
mass (g)&20.98 &20.98 &20.99 &20.92 &21.09 &21.07 &21.01 & 21.00 \\ 
LiF manufacturer&BIC&BIC&BIC&BIC&OKEN&OKEN&OKEN&OKEN \\
thermistor&NTD2&NTD3&NTD3&NTD3&NTD4&NTD3&NTD3&NTD3\\  
Etching ($\mu$m)& 21 &22&26&29&22&24&27&30 \\ \hline
\end{tabular}

\caption{LiF bolometers used for this underground measurement.}
\label{boloIDs}
\end{center}
\end{table}

The bolometer array and the inner shields are suspended with three
Kevlar cords from the bottom of the {\dil} which is cooled down to
below 10mK. Thermal connection to the {\dil} is achieved by four annealed OFHC copper ribbons of 10 mm width, 120 mm length, and 0.1 mm thickness. Each Kevlar cord is 8 cm long and eliminates the microphonic noise due to the helium liquefier. The {\dil} is constructed with the low-radioactivity materials assayed with a {\HPGe} in advance.

\begin{figure}[p]
   \begin{center}
\includegraphics[width=1.0\linewidth]{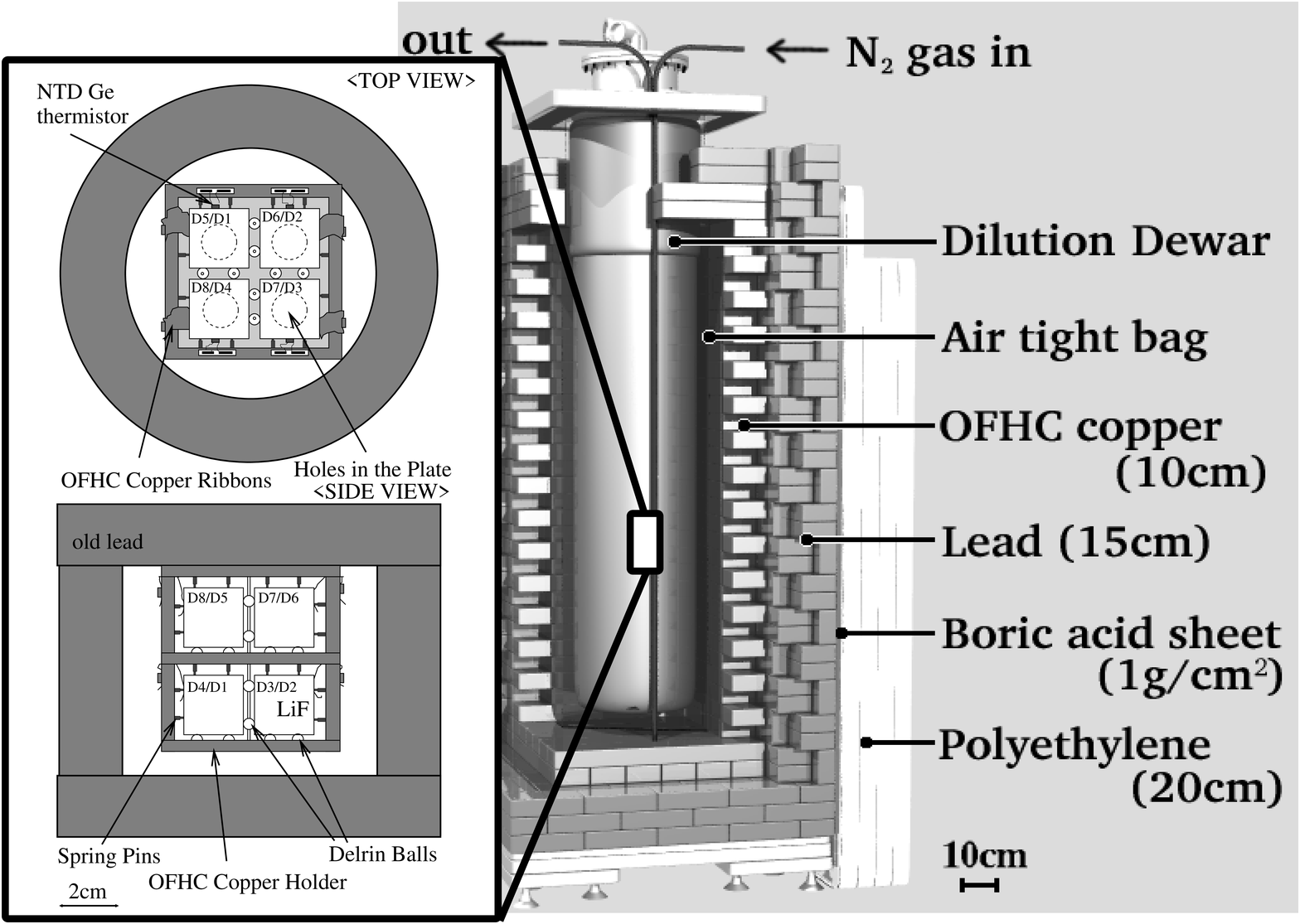}
   \caption{Schematic drawings of the bolometer array. }
   \label{sche bolometer array}
  \end{center}
   \end{figure}

The refrigerator is shielded with 10 cm of OFHC copper, 15cm of lead, $\rm 1 g/cm^2$ of boric acid sheet, and 20 cm of polyethylene as shown in Fig. \ref{sche bolometer array}. Materials for the passive shields were previously radio-assayed with a {\HPGe}\cite{Miuchi_M}. An air tight plastic bag is set between the copper shield and the cryostat. The plastic bag is filled with nitrogen gas evaporated from liquid nitrogen. 
The radon concentration was less than 0.2 Bq/$\rm m^3$ throughout the measurement period, which makes negligible background to this dark matter search experiment.

Each thermistor is DC biased through a 300 M$\rm \Omega$ load resistor. The bias current of each bolometer is listed in Table \ref{detector response}. The voltage change across the thermistor is fed to the voltage sensitive amplifier placed at the 4K cold stage.  A junction field effect transistor (J-FET), NEC 2SK163, is used for the cold stage amplifier with a gain of 10. The signal from the cold stage amplifier is in turn amplified by the voltage amplifier placed at the room temperature with a gain of 4.86 $\times10^2$. The signal from the voltage amplifier undergoes through a low pass filter with a cut-off frequency of 11 Hz and is digitized with a sampling rate of 1 kHz. All the data are recorded continuously without any online trigger for a complete offline analysis. The data size for an hour measurement is 74M bytes.

The measurement was performed from November 22, 2001 through January 12, 2002. 
Eight of the eight bolometers worked, however, D1 and D5 showed poor
signal to noise ratios because of the thermistors (NTD2 and
NTD4). We analyzed the data obtained with  the other six bolometers.  

We applied an offline level trigger to the data averaged over $\pm$10 msec. Trigger levels are determined so that the trigger rates do not exceed 1 Hz. The trigger levels and the trigger rates are shown in Table \ref{detector response}.
We determined the energy of each event by fitting the scaled template signal to the raw data.
The electric noise events and the events that are thought to have occurred within the thermistors are eliminated by the wave form analysis. Then the events above the threshold energy that is determined as the energy with efficiency=0.8 are selected as real bolometer signals. The trigger and event selection efficiencies are estimated by adding the scaled template signals to the raw data as described in Ref. \cite{Tokyo_1999_PLB}.Finally, the real bolometer events in coincidence with two or more detectors within $\pm$ 10 msec are discarded since WIMPs hardly interact in two or more detectors simultaneously.
Energy resolutions of the bolometers are estimated by the same method as the efficiency estimation.  The threshold energies used for the event selection and estimated energy resolution near the threshold are listed in Table \ref{detector response}.

The calibrations are carried out by irradiating $\rm {}^{60}Co$ and
$\rm {}^{137}Cs$ gamma-ray sources suspended between the heavy shield
and the cryogenics. Compton edges are used and good agreements with
the simulations are seen. Good linearities are seen as shown in
Ref. \cite{Tokyo_1999_PLB}. Linearities down to 60keV are confirmed
previously\cite{Tokyo surface}.  Recorded pulse heights to a given
signal for six detectors are shown as 'gain' in Table \ref{detector
response}. Gains are within the range of factor 7 which seems to be
reasonable, considering that the assembly conditions are not exactly the same for the six bolometers. Stability of the detector response and linearities are checked every three or four days and the gains are found to be stable within $\pm$ 5$\%$.

\begin{table}[h]
\begin{center}
\begin{tabular}{ l|cccc cc}
\hline
Detector&D2&D3&D4&D6&D7&D8 \\ \hline
Bias Current[nA]&0.05&0.04&0.05&0.05&0.09&0.07 \\ 
Gain[mV/keV]&0.62&1.1&2.1&1.2&0.38&2.5 \\
Trigger level[mV]&8&8&15&6&3&5.5 \\ 
Trigger rate [Hz]& 0.15&0.12&0.57&0.011&0.35&0.14  \\ 
Threshold [keV]&44.1&24.1&60.8&15.7&46.8&8.0  \\
Event rate[$10^{-3}$Hz]&0.71&1.2&1.0&0.77&0.51 &1.4 \\
Resolution (FWHM)[keV]&13&8.0&10& 3.5&7.8&1.9  \\ 
Live Time[days] &33.2&33.4&30.1&34.3&31.7&33.3 \\  \hline
\end{tabular}
\caption{Detector responses of the six bolometers. }
\label{detector response}
\end{center}
\end{table}

\section{Results}
\label{result}
\begin{figure}[p]
   \begin{center}
\includegraphics[width=1.0\linewidth]{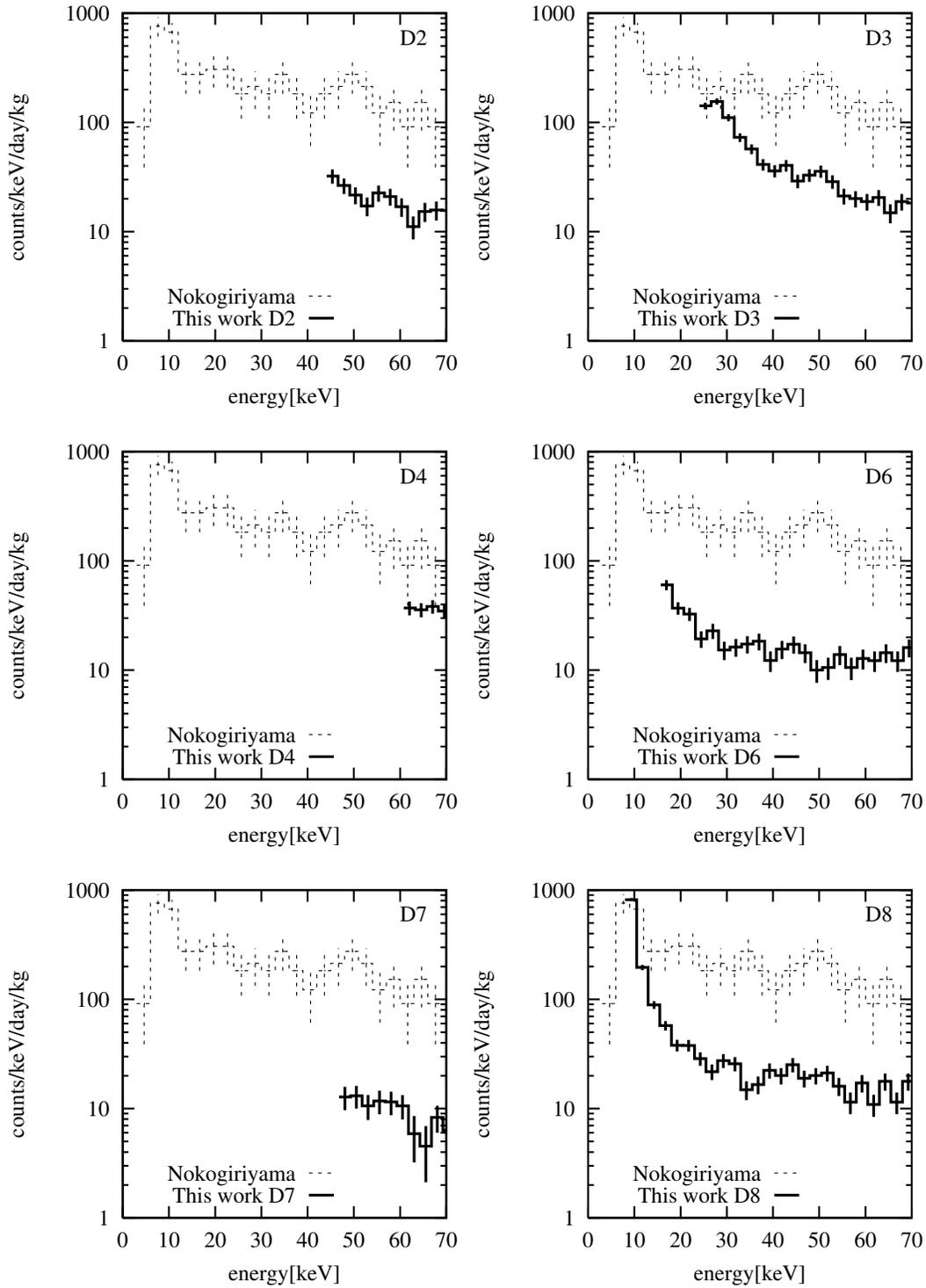}
   \caption{Low energy spectra obtained with the six bolometers are shown in thick solid lines. One of the spectra (D3) obtained in the pilot run at {\noko}\cite{Tokyo_1999_PLB} is shown in dashed lines.}
   \label{fig all spectra low}
  \end{center}
   \end{figure}

The spectra obtained with the six bolometers are shown in Fig. \ref{fig all spectra low}. One of the spectra (D3) obtained in the pilot run at {\noko}\cite{Tokyo_1999_PLB} is shown for comparison. Live times of the six bolometers are listed in Table \ref{detector response}. The total exposure is 4.1 kg$\cdot$days. Background levels of D6, D7, and D8 have decreased by nearly  one order of magnitude compared to the pilot run. D4 suffers from electronics noises, therefore the background level of D4 is higher than those of other detectors. 

\begin{figure}[p]
   \begin{center}
\includegraphics[width=1.\linewidth]{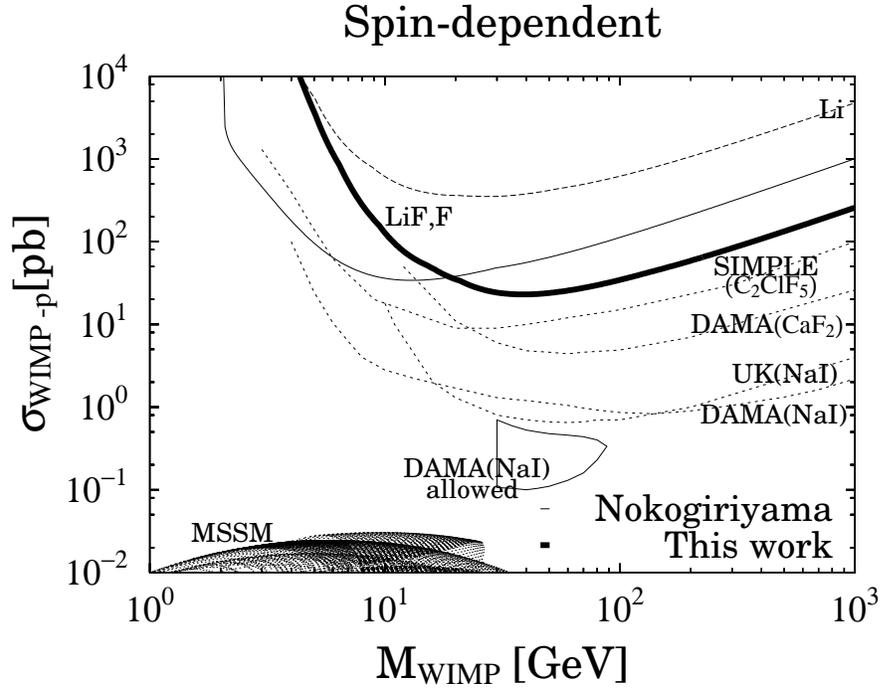}
   \caption{LiF combined 90\% C.L. $\rm \sigma^{\rm SD}_{\rm WIMP-p}$ limits as a function of $M_{\rm WIMP}$ are shown in a thick solid line.  Limits derived from Li and F alone are shown in dashed lines. 
(Limits from F are actually not seen because they almost coincide 
with the LiF combined limits shown in the thick solid line.) Limits from the pilot run and from other experiments\cite{DAMA_SD,UK_2000,DAMA_CaF2,SIMPLE} are shown in a thin solid and dotted lines, respectively.  DAMA's allowed region investigated in a mixed coupling framework is shown as ``case c'' in Ref. \cite{DAMA_SDSI}  is shown in a thin solid line. The MSSM predictions are also shown.}
   \label{fig result lim SD}
  \end{center}
   \end{figure}

From the obtained spectra, we derived the spin-dependent WIMP-proton
elastic scattering cross section ($\sigma_{\rm WIMP-p}^{\rm SD}$)
limits. The limits are shown in Fig. \ref{fig result lim SD}.We calculated the limits in the same manner as used in Ref. \cite{Tokyo_1999_PLB} and Ref. \cite{UK_1996}.
Here we used the astrophysical and nuclear parameters listed in Table \ref{table values for limit}. We used the earth velocity which corresponds to the velocity in the measurement period. We used the $\lambda^2J(J+1)$ values calculated by assuming the odd group shell model\cite{Ellis93}. For the WIMPs heavier than 50 GeV, the limits have been improved by factor 5 from the pilot run. Since we determined the energy threshold (efficiency=0.8) higher than the pilot run, we did not improve the cross section limits for the lighter WIMPs compared to the pilot run. The best limit is 23pb for $M_{\rm WIMP}=$40GeV.

\begin{table}[h]
\begin{center}
\begin{tabular}{l l}
\hline
Dark matter density& $\rho_{\rm D}=\rm 0.3GeVc^{-2}cm^{-3}$ \\ 
Velocity distribution&Maxwellian \\ 
Maxwellian velocity parameter&$v_0$=220$\rm kms^{-1}$ \\ 
Escape velocity &$v_{\rm esc}$ =650$\rm kms^{-1}$ \\ 
Earth velocity &$v_{\rm E}$=217$\rm kms^{-1}$ \\ \hline
spin factor($\rm {}^{7}Li$)&$\lambda^2J(J+1)$=0.411\\ 
spin factor($\rm {}^{19}F$) &$\lambda^2J(J+1)$=0.647\\ \hline
\end{tabular}
\caption{Astrophysical and  nuclear parameters used to derive the exclusion limit.}
\label{table values for limit}
\end{center}
\end{table}

We also derived the spin-dependent WIMP-proton and  WIMP-neutron elastic scattering cross section limits from $\rm{}^{7}Li$ and $\rm{}^{19}F$ independently in the same manner as described in Ref. \cite{Tovey}, or using 
\begin{equation}
\sigma^{\rm SD lim}_{\rm WIMP-p(N)}=\sigma^{\rm SD lim}_{\rm WIMP-N}\frac{\mu^{2}_{\rm WIMP-p}}{\mu^{2}_{\rm WIMP-N}}\left( \frac{C^{\rm SD}_{\rm p(N)}}{C^{\rm SD}_{\rm p}}\right)^{-1} 
\label{eq C proton}
\end{equation}
and
\begin{equation}
\sigma^{\rm SD lim}_{\rm WIMP-n(N)}=\sigma^{\rm SD lim}_{\rm WIMP-N}\frac{\mu^{2}_{\rm WIMP-n}}{\mu^{2}_{\rm WIMP-N}}\left( \frac{C^{\rm SD}_{\rm n(N)}}{C^{\rm SD}_{\rm n}}\right)^{-1},
\label{eq C neutron}
\end{equation}
where $\sigma^{\rm SD lim}_{\rm WIMP-p(N)}$ and $\sigma^{\rm SD lim}_{\rm WIMP-n(N)}$ are the spin-dependent WIMP-proton and WIMP-neutron cross section limits obtained on the assumption that all events are due to WIMP-proton and WIMP-neutron elastic scatterings in the nucleus N, respectively, $\sigma^{\rm SD lim}_{\rm WIMP-N}$ is the WIMP-nucleus cross section limit which is directly set by the experiments, $\mu_{\rm WIMP-p}$ and $\mu_{\rm WIMP-n}$ are the WIMP-proton and WIMP-neutron reduced masses, respectively, $\mu_{\rm WIMP-N}$ is the WIMP-nucleus reduced mass, $C^{\rm SD}_{\rm p(N)}$ and $C^{\rm SD}_{\rm n(N)}$ are the proton and neutron contributions to the total enhancement factor of nucleus N, respectively, and  $C^{\rm SD}_{\rm p}$ and $C^{\rm SD}_{\rm n}$ are the enhancement factors of proton and neutron themselves, respectively. Ratios of the enhancement factors calculated by assuming the odd group shell model are shown in Table \ref{SpSn}. 

 \begin{table}[b]
  \begin{center}
  \begin{tabular}{c|c|cc|cc}
   \hline
  Isotopes    & $J$ & $\langle S_{\rm p(N)}\rangle $&$\langle S_{\rm n(N)}\rangle $&$C^{\rm SD}_{\rm p(N)}/C^{\rm SD}_{\rm p}$&$C^{\rm SD}_{\rm n(N)}/C^{\rm SD}_{\rm n}$ \\
 \hline
    $^7{\rm Li}$ & 3/2 & 0.497& 0.004&$\rm 5.49\times10^{-1}$&$\rm 3.56\times10^{-5}$  \\
    $^{19}{\rm F}$ & 1/2&0.441  & --0.109&$\rm 7.78\times10^{-1}$&$\rm 4.75\times10^{-2}$  \\ 
    $^{23}{\rm Na}$ & 3/2&0.248  & 0.020&$\rm 1.37\times10^{-1}$&$\rm 8.89\times10^{-4}$  \\ 
    $^{127}{\rm I}$ & 5/2&0.309  & 0.075&$\rm 1.78\times10^{-1}$&$\rm 1.05\times10^{-2}$  \\ \hline
        \end{tabular}
\caption{Values of $\langle S_{\rm p(N)}\rangle ,\langle S_{\rm n(N)}\rangle ,C^{\rm SD}_{\rm p(N)}/C^{\rm SD}_{\rm p},C^{\rm SD}_{\rm n(N)}/C^{\rm SD}_{\rm n}$ for various isotopes. Values for ${}^{7}$Li and ${}^{19}$F are taken from Ref.\cite{LiSpSn} and those for ${}^{23}$Na and  ${}^{127}$I are taken from Ref.\cite{127SpSn}}
\label{SpSn}
\end{center}
\end{table}

The enhancement factor and the cross section are related by the following equation:
\begin{eqnarray}
\sigma^{\rm SD}_{\rm WIMP-N}&=&4G_{\rm F}\mu^2_{\rm WIMP-N}C^{\rm SD}_{\rm N}\nonumber \\ 
&=&4G^2_{\rm F}\mu^2_{\rm WIMP-N}\frac{8}{\pi}(a_{\rm p}\langle S_{\rm p(N)}\rangle +a_{\rm n}\langle S_{\rm n(N)}\rangle )^2\frac{J+1}{J},
\end{eqnarray}
where $\sigma^{\rm SD}_{\rm WIMP-N}$ is the WIMP-nucleus cross section, $G_{\rm F}$ is the Fermi coupling constant, $C^{\rm SD}_{\rm N}$ is total enhancement factor of nucleus N, $a_{\rm p}$ and $a_{\rm n}$ are the WIMP-proton and WIMP-neutron couplings, respectively, $\langle S_{\rm p(N)}\rangle$ and $\langle S_{\rm n(N)}\rangle$ are the expectation values of the proton and neutron spins within the nucleus N, respectively, and $J$ is the total nuclear spin. $\langle S_{\rm p(N)}\rangle$, $\langle S_{\rm n(N)}\rangle$, and $J$ can be calculated with shell models and are shown in Table \ref{SpSn}. 


From the obtained WIMP-proton and WIMP-neutron elastic scattering cross section limits, we derived the limits in the $a_{\rm p}-a_{\rm n}$ plane. We again followed  Ref. \cite{Tovey}. The allowed region in the $a_{\rm p}-a_{\rm n}$ plane is defined as follows:
\begin{equation}
\left(\frac{a_{\rm p}}{\sqrt{\sigma^{\rm SD lim}_{\rm WIMP-p(N)}}}\pm\frac{a_{\rm n}}{\sqrt{\sigma^{\rm SD lim}_{\rm WIMP-n(N)}}}\right)^2 <\frac{\pi}{24G_{\rm F}^2\mu^2_{\rm WIMP-p}},
\end{equation}
where the relative sign inside the square is determined by the sign of  $\langle S_{\rm p(N)}\rangle/\langle S_{\rm n(N)}\rangle$.

\begin{figure}[p]
   \begin{center}
\includegraphics[width=0.7\linewidth]{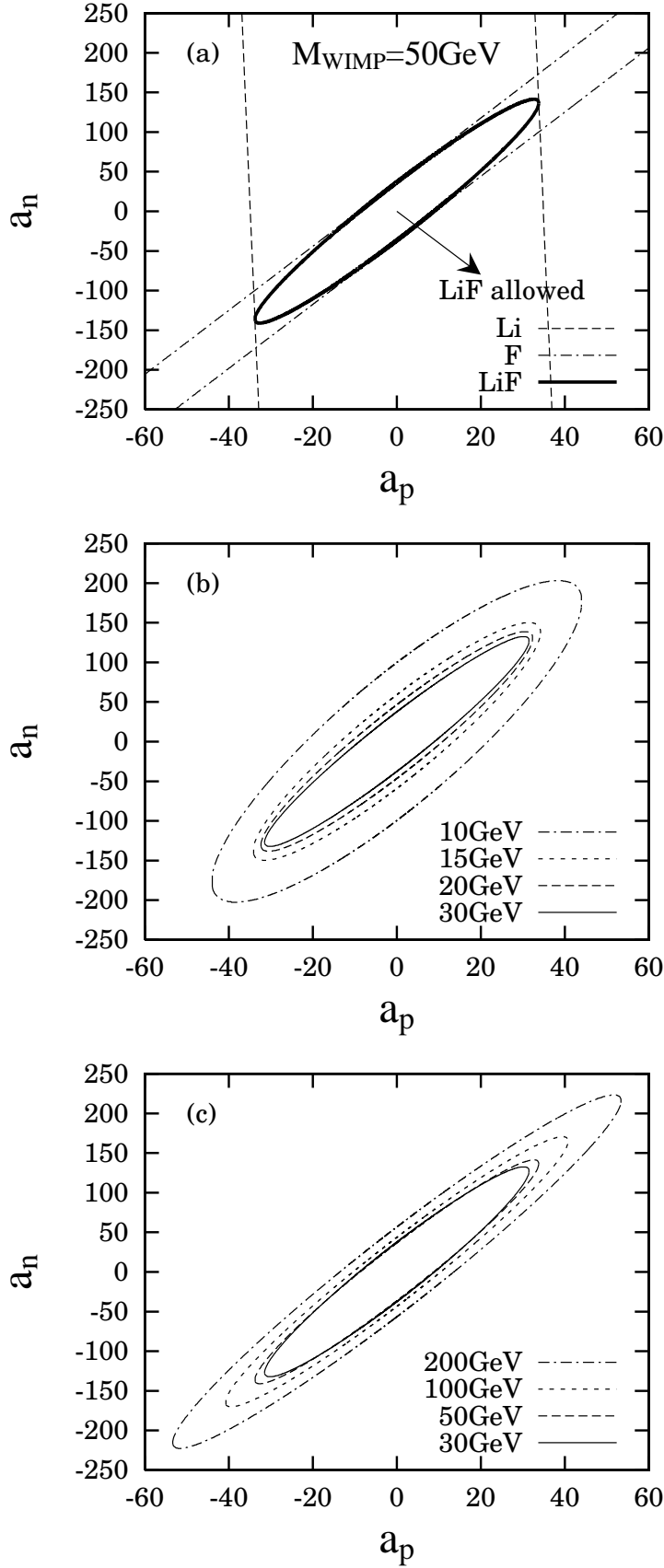}
	\caption{Limits in the $a_{\rm p}-a_{\rm n}$ plane.: (a) For $\rm M_{\rm WIMP}=50GeV$, limits set by Li(F) alone are shown in  dashed(dash-dotted) lines while the combined limits are shown in solid lines. (b) Limits for the light WIMPs. (c) Limits for the heavy WIMPs.}
   \label{fig result apan}
  \end{center}
   \end{figure}

Obtained limits in the $a_{\rm p}-a_{\rm n}$ plane for WIMPs with mass
of 50 GeV are shown in Fig.\ref{fig result apan} (a). The regions
between the two parallel lines are the allowed regions from ${}^{7}$Li
and ${}^{19}$F alone, respectively. It is seen that a single nuclide is not sufficient to set a finite limit in the $a_{\rm p}-a_{\rm n}$ plane. Since we have ${}^{7}$Li and ${}^{19}$F as the targets, we can take the intersection of the two regions as the combined allowed region from the LiF detector. This is the merit of the detector consisting of more than two nuclides with sufficient spins. We show the allowed regions in the $a_{\rm p}-a_{\rm n}$ plane obtained by this experiment for light WIMPs (Fig.\ref{fig result apan} (b)) and heavy WIMPs (Fig.\ref{fig result apan} (c)). The limits set by this experiment for $M_{\rm WIMP}$=30GeV are $|a_{\rm p}|<32$ and $|a_{\rm p}|<133$. 

We compared our results with those of other experiments. As the
DAMA group had not shown the results from ${}^{23}$Na and ${}^{127}$I
independently\cite{DAMA annual,DAMA_SD,DAMA_SDSI}, we were not able to
compare our results with the spin-dependent annual modulation allowed
regions which is shown in Fig. \ref{fig result lim SD}\cite{DAMA_SDSI}. Instead, we compared our
results with those of the UKDMC
experiments\cite{UK_1996,UK_2000}. Since UKDMC shows the limits from
${}^{23}$Na and ${}^{127}$I independently \cite{UK_1996}, we can
calculate the limits in the $a_{\rm p}-a_{\rm n}$ plane from
${}^{23}$Na and ${}^{127}$I and combine them  without any
assumptions. 
It should be noted that limits in the $a_{\rm p}-a_{\rm n}$ plane from the NaI
detector are largely affected by the sensitivity ratio of ${}^{23}$Na
and ${}^{127}$I; the threshold of the detector and the quenching
factor of ${}^{127}$I largely change the limits. Furthermore, accurate calculation of the annual modulation allowed region requires all of the obtained experimental data with time information, which we do not have.
Therefore, we did not take the risk of deriving the limits or allowed region in the $a_{\rm p}-a_{\rm n}$ plane from the experiments of which independent results from ${}^{23}$Na and ${}^{127}$I are not shown. 

\begin{figure}[p]
   \begin{center}
\includegraphics[width=0.7\linewidth]{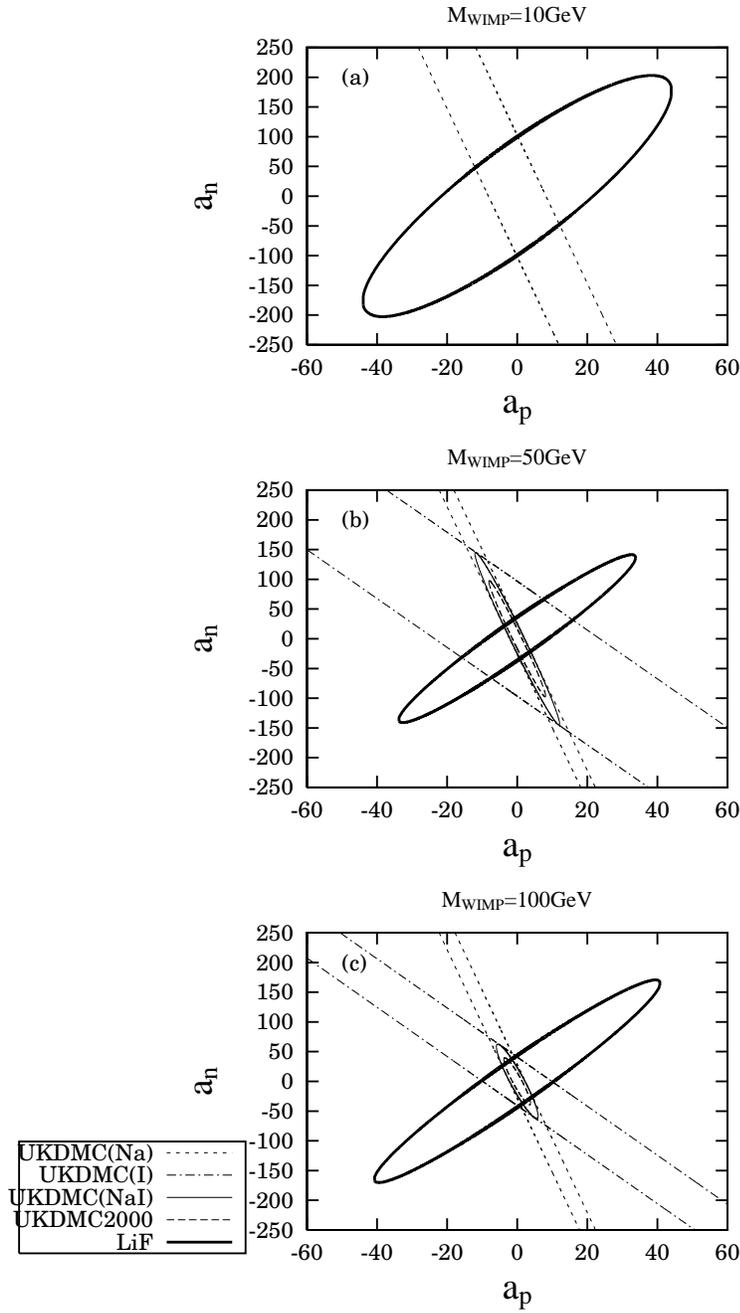}
	\caption{Comparison of the limits in the $a_{\rm p}-a_{\rm n}$ plane with the UKDMC data\cite{UK_1996} for various WIMPs masses.}
   \label{fig apan with uk}
  \end{center}
   \end{figure}

In Fig. \ref{fig apan with uk}, the allowed regions for the (a):
sufficiently light WIMPs for which  ${}^{127}$I  loses the sensitivity
and NaI effectively becomes a detector with one nuclide, (b): WIMPs
with mass near the center value of DAMA's annual modulation signal,
and (c): WIMPs heavy enough for which ${}^{23}$Na and ${}^{127}$I both
have good sensitivities, are shown. In the three plots, the limits
from Na and I of UKDMC\cite{UK_1996} are shown in dotted and
dash-dotted lines, respectively, NaI combined limits of
UKDMC\cite{UK_1996} are shown in thin solid lines and LiF combined
limits from this work are shown in thick solid lines. UKDMC published
the recalculated data\cite{UK_2000} in which only NaI combined limits are shown, and we derived the limits in the $a_{\rm p}-a_{\rm n}$ plane
on the assumption that the sensitivity ratios of ${}^{23}$Na and ${}^{127}$I are the same as those shown in Ref. \cite{UK_1996}. These new limits from Ref. \cite{UK_2000} are shown in dashed lines just for reference.

It is seen that our results can set limits complementary to the
results obtained with NaI detectors. For the light WIMPs (lighter than
20 GeV for most of the detectors) to which ${}^{127}$I  loses
sensitivity, this is the only experiment that can set a finite allowed
region in the $a_{\rm p}-a_{\rm n}$ plane. 

For the WIMPs with a mass
of 50 GeV, which is near the center value of DAMA's annual modulation signal, we exclude more than two thirds of the parameter space
allowed by the UKDMC experiments. As the limits from the UKDMC experiments are close to the DAMA's spin-dependent allowed region, as shown in Fig. \ref{fig result lim SD}, we should also be able to  compare our limits in the $a_{\rm p}-a_{\rm n}$ plane.  We, however, have to wait until DAMA publishes relevant data to calculate the allowed region in the $a_{\rm p}-a_{\rm n}$ plane.

 For the WIMPs with a mass of 100 GeV, we only exclude a part of the UKDMC allowed region.  However, it must be stressed that though our $\sigma_{\rm WIMP-p}^{\rm SD}$ limits are about two orders of magnitude worse than the limits of UKDMC as shown in Fig. \ref{fig result lim SD} , we are able to tighten their limits in the $a_{\rm p}-a_{\rm n}$ plane. This is because  the spin factors of ${}^{19}$F are  large and the $\langle S_{\rm p(N)}\rangle/\langle S_{\rm n(N)}\rangle$ sign of ${}^{19}$F is different from those of ${}^{7}$Li, ${}^{23}$Na, and ${}^{127}$I.


\section{Discussions and prospects}
\label{discussions}
Though we are able to tighten the limits from the UKDMC experiment in the  $a_{\rm p}-a_{\rm n}$ plane, the background is still high for the direct detection of WIMPs.
We studied the background sources and found that potassium contamination on the surface of the LiF crystals, tritium produced by the neutron capture of $\rm {}^{6}Li$ during the development in the surface laboratory, and the uranium and  thorium contamination in the OFHC copper holder and in the inner lead shield dominate the background in the energy region between 10 keV and 90 keV. 
We are planning to etch the LiF crystals more deeply in order to eliminate the contamination on the surface of the LiF crystals. We also plan to use NaF crystals as a bolometer in order to eliminate the effects of tritium. It is also expected that NaF crystals give more stringent limits in the $a_{\rm p}-a_{\rm n}$ plane since  $^{19}{\rm F}$ and $^{23}{\rm Na}$ set limits orthogonal to each other in the $a_{\rm p}-a_{\rm n}$ plane as shown in Fig. \ref{fig result apan} and Fig. \ref{fig apan with uk}.
We are planning to install active shields between the LiF crystals and the OFHC copper holders to reduce the background from the OFHC copper holder and the inner shield. With these improvements, we expect to improve the sensitivity of our detector by more than one order of magnitude.

\section{Conclusions}
\label{Conclusions}
We have performed the first underground dark matter experiment at
{\kam}. With a total exposure of 4.1 kg$\cdot$days, we have obtained
$\sigma_{\rm WIMP-p}^{\rm SD}$ limits factor 5 improved from the pilot
run at {\noko}. We calculated the limits in the $a_{\rm p}-a_{\rm n}$
plane and showed that our results tightened the limits from the UKDMC
experiment. For the WIMPs with a mass of 50 GeV, we have excluded more than two thirds of the parameter space in the $a_{\rm p}-a_{\rm n}$ plane allowed by the UKDMC experiment. As DAMA's spin-dependent annual modulation allowed region in Fig. \ref{fig result lim SD} is close to UKDMC's limits, we should also be able to  compare our limits in the $a_{\rm p}-a_{\rm n}$ plane with DAMA's spin-dependent annual modulation allowed region when DAMA publishes relevant data to calculate the allowed region in the $a_{\rm p}-a_{\rm n}$ plane.



\section*{Acknowledgements}
We would like to thank all the staffs of Kamioka Observatory of the
Institute for Cosmic Ray Research, the University of Tokyo, to whose
hospitality we owe a great deal in using the facilities of the
Observatory. We would also like to express our gratitude to Prof. Komura at Kanazawa University for providing us with 25kg of low-activity old lead. This research is supported by the Grant-in-Aid for COE Research by the Japanese Ministry of Education, Culture, Sports, Science and Technology.

\newpage
\pagestyle{empty}

\newpage
\newpage

\end{document}